\documentstyle[prl,aps,multicol,epsf]{revtex}

\tighten
\begin{document}

\draft
\title{Density of states  of a two-dimensional electron gas in a non-quantizing magnetic
field}
\author{A.M. Rudin$\dagger$, I.L. Aleiner$^*$, and L.I. Glazman}
\address{Theoretical Physics Institute, University of Minnesota, Minneapolis MN
55455}
\maketitle

\begin{abstract}
We study  local density of electron states of a
two-dimentional  conductor with a smooth disorder potential in a non-quantizing magnetic
field, which does not cause the standart de Haas-van Alphen oscillations. It is found,  that
despite the influence of  such ``classical''  magnetic field on the average
electron density of states (DOS) is negligibly small,  it  does produce a significant effect
on the DOS correlations. The corresponding correlation function exhibits oscillations  with
the characteristic period  of cyclotron quantum $\hbar\omega_c$.
\end{abstract}

\begin{multicols}{2} 

\section{Introduction}

In the clean homogeneous electron gas the wave functions of electrons are plane waves and
the density of electron gas is constant in space.  In disordered conductors electrons are
scattered by impurities, which change their wave functions from the plane waves. This, in
turn, results in spatial variations of the electron density.  It is
appropriate to describe those variations by introducing the  local density of electron states,
$\nu(\epsilon, {\bf r})$, which is determined by the  equation
\begin{equation} \label{def}
\nu(\epsilon, {\bf r}) =  2\sum_{\alpha} \left| \psi_\alpha
({\bf r}) \right|^2 \delta (\epsilon - \epsilon_\alpha ),
\end{equation}
where index $\alpha$ specifies the electronic states, and the factor of two reflects the
spin degeneracy. 
The distribution function of the local DOS at the fixed energy in  open metallic disordered
samples was studied in many papers (see {\it e.g.} \cite{Many}) with the emphasis on the rare,
nontypical fluctuations. It was found that although the local DOS distribution is close to the
Gaussian one, it has slowly decaying logarithmically normal asymptotics. 
Prigodin\cite{Prigodin93} studied correlation function of the  density of the electron 
states of a two-dimensional system at different energies in relation to 
the NMR line shape. 

It is well-known that  strong magnetic field modifies the single-particle
density of electron states, both local and average,  due to the Landau
quantization. In a two-dimensional
electron  gas the quantization leads to a peak structure in the average density of states,
which is  revealed in tunneling experiments as peaks in the dependence of the tunneling
conductance  on the applied bias, see, {\it e.g.}, Ref.
\cite{Eisenstein}. The form and  width of these peaks are
determined\cite{Ando} by the disorder. 

In a weak  magnetic field  the distance between the Landau levels,
$\hbar\omega_c$, is smaller than  their disorder-induced width. As a result, in such
``classical'' magnetic field, oscillations in the   average density of states caused by the 
Landau quantization become exponentially small \cite{Ando}, $\propto \exp[-2\pi/(\omega_c
\tau_s)]$. Here $\tau_s$ is a quantum lifetime of an electron.

 The
goal of the present paper is to show that, despite such ``classical''  magnetic field does
not influence the average DOS,  it  does produce a significant effect on the
 correlation function of the {\it local} density of states fluctuations
\begin{equation} \label{P}
P(\epsilon_1,\epsilon_2, {\bf r}) = { \left< \delta \nu (\epsilon_1, {\bf r}) \delta \nu
(\epsilon_2, {\bf r}) \right> \over \nu_0^2}.
\end{equation}
Here $\delta \nu (\epsilon, {\bf r}) = \nu (\epsilon, {\bf r}) - \nu_0$ is the
local  deviation of the DOS in point ${\bf r}$ from its average value, $\nu_0=m/\pi\hbar^2$,
$m$ is the electron mass, and brackets  $\left< \mbox{ } \right>$ denote averaging over the
random impurity potential. Clearly, the correlation function depends only on difference
of energy agruments, $P(\epsilon_1, \epsilon_2, {\bf r}) = P(\epsilon_1-\epsilon_2, {\bf
r})$. 

The effect of the classical
magnetic field on the DOS correlation function  becomes  pronounced if  the disorder
potential has a  correlation length much larger than the
Fermi wave length. In such a potential,
electrons experience small-angle scattering, and their transport relaxation time $\tau_{\rm
tr}$, is much larger than
$\tau_s$. Thus there exists a range of magnetic fields, in which Landau
quantization is suppressed ($\omega_c \tau_s \ll 1$), while classical
electron trajectories are strongly affected by the field ($\omega_c \tau_{\rm tr}\gg
1$). In this regime the correlation function, $P(\epsilon_1-\epsilon_2)$  is strongly
enhanced with respect to the zero magnetic field case and 
exhibits peaks as a function of energy difference $\epsilon_1- \epsilon_2$ with the
distance between peaks equal to the cyclotron quantum, $\hbar\omega_c$. 
For the macroscopically homogeneous sample the shape
of the $n$-th peak, $|(\epsilon_1 - \epsilon_2) - n\hbar \omega_c| \lesssim
\hbar \omega_c/2$, in the local DOS correlation function is given by:
\begin{equation}\label{result}
P(\epsilon_1-\epsilon_2) =  {\omega_c^2\tau_{tr}^2\over
2\sqrt{2}\pi E_F \tau_{tr}} \,  {1 \over n} \,
f \!\left({\epsilon_1 - \epsilon_2 -n \hbar \omega_c \over \hbar  n^2 /\tau_{\rm tr}}
\right),
\end{equation}
where
\begin{equation}\label{result1}
f(x) = {1\over \sqrt{2}} \left[{1+\sqrt{x^2+1}\over x^2+1} \right]^{1/2}, 
 \end{equation}
 and $E_F$ is the Fermi energy. As $n$ gets bigger, the width of the peaks increases and
their height decreases, so that eventually the oscillatory structure is washed out. The total
number of  resolved peaks is of the order of $\sqrt{\omega_c\tau_{\rm tr}}$. 

Sensitivity of the correlation function $P$ to the classical magnetic field comes from
the fact, that  this function is directly associated with
the self-crossing of
classical  electron trajectories. We denote the probability for an electron
to complete a loop of
self-crossing trajectory over time $t$ as $K(t)$. The correlation function, $P(\epsilon_1-
\epsilon_2)$, turns out to be proportional to the Fourier transform of this return 
probability,
$$
P(\epsilon_1- \epsilon_2)  \propto K(\epsilon_1 - \epsilon_2) =
\int_0^{\infty} dt e^{-i(\epsilon_1 - \epsilon_2) t} K(t).
$$
 The  strong enough,
$\omega_c\tau_{\rm tr}
\gg 1$, magnetic field curves the electron trajectories,  significantly
affects the return
probability and, in turn, leads to  specific  correlations in the local DOS
at $\epsilon_1-\epsilon_2 \approx \hbar \omega_c$.

For long time scales $t \gg  \tau_{\rm tr}$, the  function $K(t)$ can be found
from the diffusion
equation. It gives $K(t)
\propto (Dt)^{-1}$ for the two-dimensional case ($D$ is the diffusion
coefficient).  The Fourier transform, $K(\omega)$, is proportional to  $
\ln(\omega)$, which leads  to the  well-known\cite{Prigodin93} logarithmic form of the
local DOS correlation function  with the renormalized by the magnetic field
diffusion coefficient.

At short time scales, $t\ll
\tau_{\rm tr}$, electrons move ballistically along the cyclotron orbits.
Provided that
$\omega_c\tau_{\rm tr} \gg 1$, during the time
$t$ electron may return to the initial point many times.  Multiple
periodic returns of electron  produce  peaks in the probability Fourier
transform
$K(\omega)$ at energies, which are multiples of the cyclotron quantum.
Correlation function $P(\epsilon_1-\epsilon_2)$ oscillates with the same period, which is
reflected by Eq. (\ref{result}).

\section{Derivation of the DOS correlation function}

Now we derive expression for the correlation function of the local DOS, $P$, 
valid for arbitrary electron energies. We omit the Planck constant in all the intermediate
formulas.  The DOS, Eq. (\ref{def}),  can be rewritten in
terms of the exact retarded and  advanced Green's functions of an electron in the
following way:
\begin{equation} \label{def2}
\nu(\epsilon, {\bf r})  =  {1 \over  \pi i} \left[ 
{\cal G}^A_\epsilon ({\bf r}, {\bf r}) - {\cal G}^R_{\epsilon} ( {\bf r}, {\bf r}) \right],
\end{equation}
where 
\begin{equation} \label{Grdef}
{\cal G}^R_{\epsilon} ( {\bf r}_1, {\bf r}_2) = \sum_\alpha {\psi_\alpha^*({\bf r}_1)
\psi_\alpha({\bf r}_2) \over \epsilon - \epsilon_\alpha + i0}, 
 \end{equation}
and ${\cal G}^A_{\epsilon} ( {\bf r}_1, {\bf r}_2) = [{\cal G}^R_{\epsilon}( {\bf r}_2,
{\bf r}_1)]^*$. Single electron wave functions $\psi_\alpha ({\bf r})$ satisfy the
Schr\"odinger equation  for noninteracting electrons, $\hat H_0 \psi_\alpha =
(\epsilon_\alpha+E_F)\psi_\alpha$, where $\hat H_0 = -(\hbar^2 / 2 m ) \nabla^2 + U_{
r}({\bf
r})$, and $U_{r}({\bf r})$ is the random potential.   

With the help of Eq. (\ref{def2}), the DOS correlation function, Eq. (\ref{P}),  can be
rewritten in terms of the ensemble-averaged products of the electron Green's functions:
\begin{eqnarray} \label{P1}
P(\epsilon_1-\epsilon_2, {\bf r}) & = & {1\over (  \pi \nu_0)^2} \left[ 2
{\bf Re} \left< {\cal G}^R_{\epsilon_1}(  {\bf r}, {\bf r}) {\cal G}^A_{\epsilon_2}({\bf
r}, {\bf r})
\right>  \right. \nonumber \\ 
& + &  \left< {\cal G}^R_{\epsilon_1}( {\bf r}, {\bf r}) {\cal G}^R_{\epsilon_2}( {\bf r},
{\bf r})
\right>\!\! + \!\! \left.
\left< {\cal G}^A_{\epsilon_1}( {\bf r}, {\bf r}) {\cal G}^A_{\epsilon_2}( {\bf r}, {\bf
r})
\right>\right]. \nonumber \\
 & &  
\end{eqnarray}
 The averages of the  type $ \left<{\cal G}^R{\cal G}^R \right>$ and $ \left< {\cal
G}^A{\cal G}^A \right>$ can be neglected as they do not contain
contributions associated with the  electron trajectories longer than
$\lambda_F$ and, thus, do not  produce  energy dependence of $P(\epsilon_1-\epsilon_2,
{\bf r})$  at $\epsilon_1 - \epsilon_2 \ll E_F$. To the contrary,
averaged product  $ \left<{\cal G}^R{\cal G}^A \right>$  is
determined by  long electron trajectories (see e.g. Ref. \cite{aleiner}). In general,
the product of two exact Green's functions ${\cal G}^R_{\epsilon_1}({\bf r}_1,  {\bf r}_2)
{\cal G}^A_{\epsilon_2}({\bf r}_3, {\bf r}_4)$ oscillates rapidly with the distance
between its arguments, so that the function 
\begin{equation} \label{K}
K(\epsilon_1, \epsilon_2, {\bf r}_1, {\bf r}_2, {\bf r}_3, {\bf r}_4) = 
\left<{\cal G}^R_{\epsilon_1}({\bf r}_1, {\bf r}_2) {\cal G}^A_{\epsilon_2}(
{\bf r}_3, {\bf r}_4)\right>
\end{equation}
averages out. This is no longer the case if its arguments are close to each other pairwise.
Namely,  the sizes $|{\bf r}_1-{\bf r}_4|, \, |{\bf r}_2-{\bf r}_3|$ or, alternatively,
$|{\bf r}_1-{\bf r}_3|, \, |{\bf r}_2-{\bf r}_4|$ of spatial domains defining
the ends of a trajectory should be small enough (less than $v_F \tau_s$) so that electron
propagation in these two domains could be described by plane waves.

If the ends of trajectories are separated by the distance exceeding electron
wave-length,  $|{\bf r}_1-{\bf r}_2|\gtrsim \lambda_F$, one can relate the function  $K$ to
the generalized classical correlation functions -- diffuson
${\cal K}^{\cal D}$ and Cooperon ${\cal K}^{\cal C}$, which are given by the sum of all
ladder and all maximally-crossed diagrams respectively. Namely, 
\begin{eqnarray} \label{D}
K(\epsilon_1, \epsilon_2, {\bf r}_1, &{\bf r}_2, &{\bf r}_3, {\bf r}_4)\nonumber \\
 & = &
\pi\nu_0\int {d\phi_1 \over 2\pi} \int {d\phi_2 \over 2\pi}  e^{i{\bf p}_1({\bf r}_1 -
{\bf r}_4)} 
\nonumber \\
& \times & e^{i{\bf p}_2({\bf r}_3 - {\bf r}_2)}
{\cal K}^{\cal D}_{\epsilon_1-\epsilon_2 }({\bf r}_1, \phi_1; {\bf r}_2, \phi_2) \\
 |{\bf r}_1-{\bf r}_4|, |{\bf r}_2 & - & {\bf r}_3|  \ll  v_F\tau_{s} \nonumber 
\end{eqnarray}
or 
\begin{eqnarray} \label{C}
K(\epsilon_1, \epsilon_2, {\bf r}_1, &{\bf r}_2, &{\bf r}_3, {\bf r}_4) \nonumber \\
& = &
\pi\nu_0\int {d\phi_1 \over 2\pi} \int {d\phi_2 \over 2\pi}  e^{i{\bf p}_1({\bf r}_1 -
{\bf r}_3)} 
\nonumber \\
& \times & e^{i{\bf p}_2({\bf r}_4 - {\bf r}_2)}
{\cal K}^{\cal C}_{\epsilon_1-\epsilon_2 }({\bf r}_1, \phi_1; {\bf r}_2, \phi_2)\\
|{\bf r}_1-{\bf r}_3|, |{\bf r}_2 & - & {\bf r}_4|   \ll   v_F\tau_{s}.\nonumber 
\end{eqnarray}
Here ${\bf p}_i = p_F {\bf n}_i$, where ${\bf n}_i = (\cos \phi_i, \sin\phi_i)$ is a
unit vertor with the direction  determined by the angle $\phi_i$.  

In the opposite limit, when the all four points ${\bf r}_1$, ${\bf r}_2$, ${\bf r}_3$, and
${\bf r}_4$ coinside, both ladder and maximally-crossed diagrams contribute to Eq. (\ref{K}).
 As a result, the DOS correlation function, Eq. (\ref{P1}), contains both the diffuson and
the Cooperon contributions:
\begin{equation} \label{P2}
P(\epsilon_1, \epsilon_2, {\bf r}) = {2 \over   \pi \nu_0} \, {\bf Re} \left[ {\cal
D}_{\epsilon_1-\epsilon_2 }({\bf r}, {\bf r}) + {\cal C}_{\epsilon_1-\epsilon_2}({\bf r},
{\bf r})\right].  
\end{equation}
Here ${\cal D}$ and ${\cal C}$ are the diffuson, ${\cal K}^{\cal D}$ and the Cooperon,
${\cal K}^{\cal C}$, averaged over the initial and the final directions of the electron
momentum:
\begin{eqnarray} \label{Dav}
{\cal D}_{\epsilon_1-\epsilon_2}({\bf r}_1, {\bf r}_2) & = & \int {d \phi_1 \over 2
\pi} {d \phi_2 \over 2 \pi} {\cal  K}^{\cal D}_{\epsilon_1-\epsilon_2 }({\bf r}_1, \phi_1;
{\bf r}_2, \phi_2), \\
\label{Cav}
{\cal C}_{\epsilon_1-\epsilon_2}({\bf r}_1, {\bf r}_2) & = & \int {d \phi_1 \over 2
\pi} {d \phi_2 \over 2 \pi} {\cal  K}^{\cal C}_{\epsilon_1-\epsilon_2 }({\bf r}_1, \phi_1;
{\bf r}_2, \phi_2).
\end{eqnarray}

As one sees, calculation  of the  DOS correlation function reduces to the analysis of
two classical correlation functions, ${\cal D}$ and ${\cal C}$.  Provided that we are
interested in the DOS correlation function in the presence of the magnetic field, the problem
can be further simplified. Indeed,  as it is well-known, the diffusion and the Cooperon depend
quite differently on the magnetic field (see {\it e. g.} Ref. \cite{Aronov88}). In particular,
${\cal C}$ is exponentially suppressed if the magnetic length,
$\lambda_H = \sqrt{c\hbar/eH}$, becomes smaller than  the transport relaxation length
$l_{tr}$. We, in fact, assumed a much stronger condition, $\omega_c \tau_{tr} \gg 1$, for
the magnetic field. Thus the Cooperon term in Eq. (\ref{P2})  can be neglected in our
case. On the other hand,  the diffusion term in Eq. (\ref{P2}) is meaningful and will be
analyzed below. 

\section{DOS correlation function for an infinite two-dimensional electron gas}

Let us first calculate the  local DOS  correlation function, $P(\epsilon_1-
\epsilon_2)$, in  the macroscopically homogeneous sample. The generalized diffuson, ${\cal K}^{\cal D}_{\omega}({\bf r}_1,
\phi_1; {\bf r}_2, \phi_2)$,
satisfies  the Boltzmann equation (see e.g. Ref.\cite{aleiner}) describing the  scattering of
electrons on impurities in the presence of the magnetic field.  In the special case
$\tau_s \ll \tau_{\rm tr}$ we are interested in, small angles scattering   dominates the
collision integral. With account for this simplification,  the transport
equation for ${\cal K}^{\cal D}_{\omega}({\bf r}_1, \phi_1; {\bf r}_2,
\phi_2)$ takes the Fokker-Planck form:
\begin{eqnarray} \label{Boltzmann}
\Big[-i \omega & +&    v_F {\bf n}_2  {\partial \over
\partial {\bf r}_2} + \omega_c {\partial \over \partial  \phi_2} -
{1 \over \tau_{\rm tr}} {\partial^2  \over \partial \phi_2^2}\Big]
{\cal K}^{\cal D}_{\omega}({\bf r}_1, \phi_1; {\bf r}_2, \phi_2)\nonumber \\
&=& 2\pi \delta(\phi_1-\phi_2) \, \delta ({\bf r}_1 - {\bf r}_2).
\end{eqnarray}
Equation (\ref{Boltzmann}) describes electron motion
along the cyclotron orbit accompanied by the angular diffusion caused by
scattering on a random potential. The solution of this equation will give us the Fourier
transform, 
$$
{\cal K}^{\cal D}_{\omega}(1; 2) = \int_0^\infty {\cal K}^{\cal D}(t, 1; 2) e^{-i\omega t}
dt
$$
of probability density, ${\cal K}^{\cal D}(t,{\bf r}_1, \phi_1; {\bf r}_2, \phi_2)$, for
electron which starts at moment $t=0$ in point ${\bf r}_1$ with the direction of momentum
$\phi_1$ to arrive at moment $t$ to the point ${\bf r}_2$ with momentum
direction $\phi_2$.

In order to solve Eq. (\ref{Boltzmann}) it is convenient to introduce new spatial variables
which correspond to the center of the  electron cyclotron orbit:
\begin{equation}\label{rtoR}
{\bf R}  = {\bf r} + R_c [{\bf n} \times {\bf z}].
\end{equation}
Here $R_c$ is a cyclotron radius, and ${\bf z}$ is a unit vector parallel to the magnetic
field. Changing variables in Eq.~(\ref{Boltzmann}), and performing
the Fourier  transformation from ${\bf R}_2$ to ${\bf q}$, we obtain:
\begin{eqnarray} \label{Boltz1}
\Biggl\{ &-i& \omega + \omega_c {\partial \over \partial  \phi_2} +  {R_c^2 q^2\over
2\tau_{\rm tr}} - {1\over \tau_{tr}} {\partial^2 \over \partial \phi_2^2} \nonumber \\
& + & {R_c^2 \over \tau_{\rm tr}}\left[  ({\bf n}_2 {\bf q})^2 - {q^2\over 2}-{i \over
R_c} \left( {\bf n}_2 {\bf q}{\partial \over
\partial \phi_2} + {\partial \over \partial  \phi_2} {\bf n}_2 {\bf q} \right)
\right] \Biggr\} \nonumber \\ & \times & 
{\cal K}^{\cal D}_{\omega}({\bf R}_1, \phi_1; {\bf q},
\phi_2) = 2\pi \delta(\phi_1-\phi_2) e^{-i{\bf q} {\bf R_1}}.
\end{eqnarray}  

We seek for the solution of Eq. (\ref{Boltz1}) in the following form:
\begin{equation} \label{form}
{\cal K}^{\cal D}_{\omega}({\bf R}_1, \phi_1; {\bf q}, \phi_2) = 
\sum_n e^{in\phi_2}  F_n (\omega, {\bf q}; \phi_1).
\end{equation}
Substitution of Eq.(\ref{form}) to Eq. (\ref{Boltz1}) results in a linear system of
equations for $F_n$. At small enough wave vectors,  $qR_c\ll\omega_c^2 \tau_{\rm
tr}/(|\omega| +
\omega_c)$,  terms in
the square brackets in the l.h.s. of Eq. (\ref{Boltz1}) become small,  the equations
corresponding to different $n$ become independent, and we obtain a solution  for $F_n
(\omega, {\bf q}; \phi_1)$ in the form:
\begin{equation} \label{sol2}
 F_n (\omega, {\bf q}; \phi_1) = {e^{-in\phi_1}
e^{-i{\bf q} {\bf R}_1}
\over \displaystyle  -i(\omega- n\omega_c) + { R_c^2q^2 \over 2\tau_{\rm tr}} +
{n^2\over \tau_{\rm tr}}}. 
\end{equation}
The inverse transformation of variables immediately yields now  the solution of Eq.
(\ref{Boltzmann}):
\begin{eqnarray} \label{sol1}
{\cal K}^{\cal D}_{\omega}({\bf r}_1,  \phi_1; {\bf r}_2, \phi_2) &= &\int {d{\bf q} \over
(2\pi)^2} e^{i{\bf q}({\bf r}_2 - {\bf r}_1)} \nonumber \\
& \times &  \sum_n { e^{in(\phi_2 - \phi_1)} e^{iR_c{\bf q}[({\bf n}_2 - {\bf n}_1)
\times {\bf z}]} \over
\displaystyle  -i(\omega- n\omega_c) + { R_c^2q^2 \over 2\tau_{\rm tr}} + {n^2\over
\tau_{\rm tr}}}.
\end{eqnarray}

After substitution of Eq. (\ref{sol1})  into the Eqs. (\ref{P2}) and
(\ref{Dav}), and subsequent integration over angles, we obtain the following expression
for the correlation function of the local DOS of a homogeneous two-dimensional conductor 
in the classical magnetic field:
\begin{eqnarray} \label{P3} 
P(\epsilon_1-\epsilon_2) &=& {2 \over  \pi \nu_0} \, \int {d{\bf q} \over (2\pi)^2}
{\bf Re} \, {\cal D}_{\epsilon_1 -
\epsilon_2}  ({\bf q}), 
\\
\label{Dres}
{\cal D}_{\epsilon_1 - \epsilon_2}  ({\bf q}) &= &  \sum_n
 {|J_n(qR_c)|^2 \over \displaystyle 
 -i(\epsilon_1 \! - \!\epsilon_2 \!-\! n\omega_c)\! + { R_c^2q^2 \over 2\tau_{\rm tr}} +
{n^2\over \tau_{\rm tr}}}. 
\end{eqnarray}
Here $J_n(z)$ is a Bessel function. 
At small frequencies, $\omega = \epsilon_1 - \epsilon_2 \ll \hbar/\tau_{\rm tr}$, the
$n=0$  term in Eq. (\ref{Dres})  dominates, and  
$$
{\cal D}_\omega (
{\bf q}) \approx {1\over -i\omega + R_c^2q^2/2\tau_{\rm tr}}.
$$

\narrowtext
\begin{figure}
  \epsfxsize=6cm 
\centerline{\epsfbox{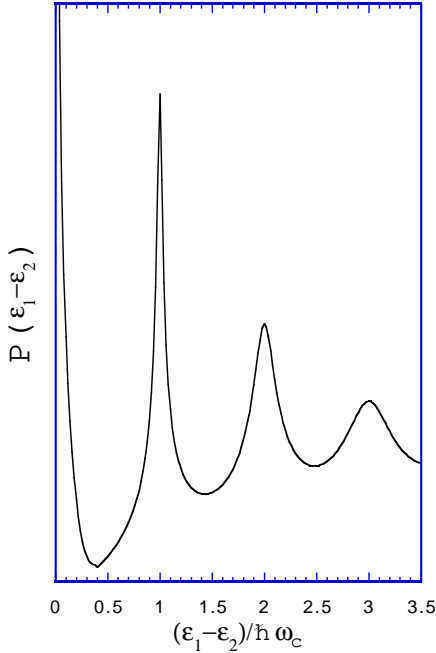}}
\caption[DOS correlation function in the classical magnetic field]
{\label{1}
Energy dependence of the local density of states correlation function, $P(\epsilon_1 -
\epsilon_2)$, of the macroscopically homogeneous sample in the classical magnetic field,
obtained by numerical analysis of Eqs. (\ref{P3})-(\ref{Dres}). Parameter $\omega_c
\tau_{tr} = 10$.}
 \end{figure}

This limit corresponds to the
diffusion regime with the diffusion coefficient $D=R_c^2/2\tau_{tr}$ renormalized by the
magnetic field. The  correlation function $P(\epsilon_1-\epsilon_2)$ depends
logarithmically on $\epsilon_1-\epsilon_2$ in this limit:
\begin{equation} \label{Pdif}
P(\epsilon_1-\epsilon_2) = {(\omega_c\tau_{tr})^2 \over 2 \pi E_F \tau_{tr}}
\ln\left[ {\hbar \over |\epsilon_1-\epsilon_2|\tau_{tr}}\right].
\end{equation}

At large frequencies, $\omega = \epsilon_1 - \epsilon_2 \gg \hbar/\tau_{\rm tr}$ the
correlation function $P(\epsilon_1-\epsilon_2)$ exhibits peaks at $\epsilon_1 -
\epsilon_2$ close to multiples of  cyclotron quantum,  $n\hbar \omega_c$.  The form of
the n-{\it th} peak is given by:
\begin{equation} \label{rescomp}
P(\epsilon_1-\epsilon_2) = { \omega_c^2\tau_{tr}^2 \over  \pi E_F \tau_{tr}} {\bf Re}
\left\{I_n[a(n, \delta \omega)]  K_n[a(n, \delta \omega)]\right\},
\end{equation}
where  $\hbar \delta \omega = \epsilon_1 -\epsilon_2-n\hbar \omega_c$ was introduced
instead of $\epsilon_1 -\epsilon_2$, and $a(n, \delta\omega)=2(n^2 - i\delta \omega
\tau_{tr})$. Functions  $I_n(a)$ and
$K_n(a)$ are the modified Bessel functions. For large $a$ we can use the 
asymptotical relation, $I_n(a)K_n(a) \approx 1/2a$, and arrive at the resulting Eq.
(\ref{result}) that describes energy dependence of the local DOS correlation function in
the vicinity of $n-$th peak.

Overall energy dependence of the correlation function of the local density of states 
for an infinite sample, obtained by numerical analysis of Eqs. (\ref{P3})-(\ref{Dres}) is
presented in Fig.~\ref{1}. The DOS correlation function exhibits strong oscillations with
the period close to $\hbar\omega_c$.

\section{Oscillations of the DOS for tunneling into the edge of a two-dimensional electron
gas}

\begin{figure}
  \epsfxsize=6cm 
\centerline{\epsfbox{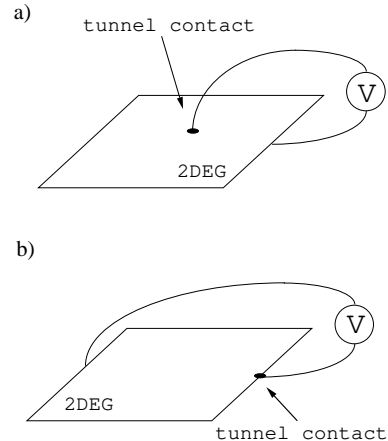}}
\caption[Measuring of the DOS correlation function]
{\label{2}
Two possible tunneling experiment that enable to measure properties of the tunneling
density of electron states: {\it (a)} the point-like tunnel contact is attached to a
two-dimensional conductor far from its edges; {\it (b)} tunneling occurs at
the edge of the two-dimensional electron gas.}
 \end{figure}

The tunneling density of states $\nu (\epsilon, {\bf r})$ is directly related to the
tunneling differential conductance $G(V)$ of a point contact attached to the two-dimensional
gas, $\nu (eV, {\bf r})/\nu_0=G(V)/G_0$ (here $G_0$ is the average linear conductance at
zero magnetic field). Thus measuring the conductance correlation function  
$\left<\delta G(V)\delta G(V+\Delta V)\right>$ one can determine the DOS correlation
function, $P(e\Delta V) = \left<\delta G(V)\delta G(V+\Delta V)\right>/G_0^2$. Here $\delta
G(V) = G(V) -  G_0$. The tunneling DOS we studied so far is related to tunneling into the
``bulk'' of a two-dimensional electron gas, see Fig.~\ref{2}a. For GaAs heterostructures,
however, there exists a well developed method of forming point contacts for lateral
tunneling into the edge of a two-dimensional electron gas (see, {\sl e.g.}, the review
of Beenakker and van Houten, Ref.\cite{Bee91}). The edge affects electron trajectories and
thus alters the correlation function of the tunneling density of states. Below we estimate
$P(\epsilon_1-\epsilon_2)$ for the specific case of lateral tunneling, schematically shown
in Fig.~\ref{2}b. We demonstrate that the oscillatory pattern of $P$ at energies larger
than  $\hbar\omega_c$ persists, although the amplitude of oscillations becomes smaller than
in  the case of tunneling into the bulk.

In order to find the conductance correlation function, one should, according to
Eq. (\ref{P2}), find the Fourier transform of the return probability, ${\cal D}(t)$, for an
electron emitted from the contact right at the edge of the electron gas.

\begin{figure}
  \epsfxsize=9cm 
\centerline{\epsfbox{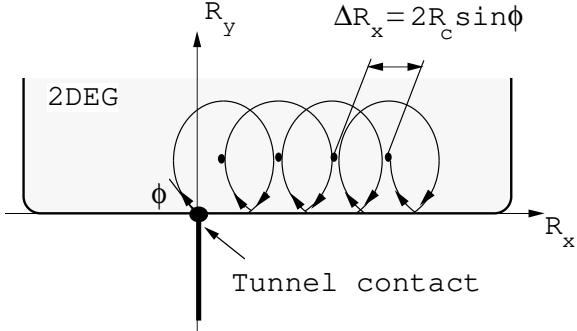}}
\caption[Drift of the electron in the magnetic field caused by the boundary scattering.]
{\label{3}
Drift of an electron in the magnetic field caused by the multiple  specular reflections
from the boundary of the two-dimensional electron gas. }
 \end{figure}

Let us consider an electron which is emitted from a point-like tunnel contact attached to the
edge of the two-dimensional conductor  at a moment
$t=0$ with the initial velocity characterized by the angle
$\phi_1$.  For any nonzero
 $\phi_1$ electron experiences multiple  reflections from the boundary
of the two-dimensional electron gas (see  Fig. \ref{3}). The boundary of the electron gas is 
usually smooth, so that we assume this scattering to be purely specular. Those multiple
scattering events lead to a drift of the guiding center of electron orbit along the boundary
of the 2DEG 
 with the velocity $v_d$, which is given by:
\begin{equation}\label{vd}
v_d = {\Delta R_x \omega_c \over 2 \pi} \approx {\sqrt{2} \over \pi} v_F \sqrt{ R_c - R_y
\over R_c} 
\end{equation}
for  $R_y$ less than $R_c$ and is zero otherwise. Here $R_x$ and $R_y$ are the 
coordinate of the center of electron orbit, which in the initial moment $t=0$ are:
\begin{equation}\label{inc}
R_x = R_c \sin \phi_1, \, \, \, R_y = R_c \cos \phi_1,
\end{equation}
and $\Delta R_x$ is defined in Fig. \ref{3}
 In the absence of disorder,  drift prevents electron from return to the
contact, and the return probability, ${\cal D}(t)=0$ at $t>0$. 
Disorder, however, makes the return probability nonzero. In fact, disorder leads to two
effects: (1) motion along the cyclotron orbit is accompanied by the angular diffusion, and
(2) in addition to  the  boundary-induced drift,  
 the guiding center of the electron cyclotron orbit diffuses in the direction
perpendicular to the boundary. As we will see, for small enough
initial angles $\phi_1$ these two effects  can, in fact,  overcome the boundary-induced drift
of electron away from the contact. 

From Eqs. (\ref{vd}) and (\ref{inc}) we see that the larger initial angle $\phi_1$ is, the
faster electron drifts away from the contact. In the view of this fact, let us start from the
case $\phi_1=0$, which correspond to the center of electron cyclotron orbit having the
initial coordinates $R_x=0$, $R_y=R_c$. Our goal now is  to obtain probability to find the
center of orbit  again in the same point after time $t$. During time
$t$ center of orbit diffuses in vertical (see Fig. \ref{3}) direction on a distance
\begin{equation}\label{dy}
\Delta Y (t)  = |R_y(t)-R_c| \approx \sqrt{Dt},
\end{equation}
 where
$D=R_c^2 /2\tau_{tr}$ is a diffusion coefficient. During the same time interval $t$,  
the center of orbit will travel along   horisontal axis on a distance
\begin{equation}\label{dx}
\Delta X = \int_0^t v_d(t') dt'\sim  v_F \int_0^t \sqrt{\Delta Y (t') \over R_c}
dt'
\approx   v_F {t^{5/4}\over \tau_{tr}^{1/4}}.
\end{equation}
Here we exploited Eqs.(\ref{vd}) and (\ref{dy}). One sees that the probability to find the   
 center of electron orbit  in the initial point after time $t$ decreases rapidly with time, 
$$
K(t) \approx {1 \over \Delta X \Delta Y} \propto {\tau_{tr}^{3/4} \over R_c v_F t^{7/4}}.
$$
As the result, in the presence of  a boundary,  contributions to
the electron return probability coming from the trajectories with two 
and more  revolutions along the cyclotron orbit are small and can be neglected, while
the main contribution  comes from trajectories which involve only one revolution
between start at $t=0$ and finish at the moment
$t\approx t_c=2\pi/\omega_c$. Let us study now this latter contribution. 

If there were no boundary, the  probability density ${\cal D}(t)$ for the electron to return
 to the initial point at time $t=t_c+\delta t$, where $\delta t \ll t_c$, could be easily
obtained from the solution, Eq. (\ref{sol1}), of the transport equation (\ref{Boltz1}). 
In fact, one  puts ${\bf r}_1 = {\bf r}_2$ in Eq. (\ref{sol1}), and integrate it 
over all possible values of $\phi_1$ and
$\phi_2$ taking into account that we are interested in the trajectories which are close to
a single cyclotron loop. As the result we obtain the return probability density which has a
strong maximum at
$t= t_c$ with the amplitude depending on the  amount of disorder in the system:
\begin{equation}\label{ddt}
{\cal D} (t\approx t_c) = { \sqrt{2} \omega_c \tau_{tr} \over R_c^2}
\exp \left[ -{ \pi \over 2} \omega_c \tau_{tr} \left( { t - t_c \over t_c}\right)^2\right].
\end{equation}
This equation is valid in the absence of the boundary, {\it i.e.} for a homogeneous
system. Clearly, for such  system trajectories with different initial
angles $\phi_1$ contribute equally to the Eq. (\ref{ddt}).  For
the system with the boundary this is obviously not the case. Namely, only a small fraction of
trajectrories with
$\phi_1
\lesssim 1/\sqrt{\omega_c \tau_{tr}}$  contribute, for which the disorder-induced uncertainty
of electron position exceeds the shift  $\Delta R_x = v_d t_c$, see  Fig. \ref{3}. Thus in the
presence of the boundary the return probability density, ${\cal D}_b(t)$, can be estimated
 by multiplying   Eq. (\ref{ddt}) by a small factor $1/\sqrt{\omega_c \tau_{tr}}$:
\begin{equation} \label{ddtb}
 {\cal D}_b(t) \approx {{\cal D} (t) \over \sqrt{\omega_c \tau_{tr}}}.
\end{equation}
According to Eqs. (\ref{P2}), the  correlation function of the DOS at the edge of the
two-dimensional electron gas, 
$P(\epsilon_1- \epsilon_2)$, is determined by the Fourier transform of ${\cal D}_b(t)$ given
by Eqs. (\ref{ddt})-(\ref{ddtb}). Performing the Fourier transformation, we finally obtain: 
\begin{eqnarray} \label{result22}
P(\epsilon_1- \epsilon_2) &\approx &{\hbar \over m \omega_c R_c^2} 
\cos  \left(2\pi {\epsilon_1- \epsilon_2 \over \hbar \omega_c} \right) \nonumber \\
  &\times &\exp \left[-{2 \pi \over \omega_c \tau_{tr}}
\left({\epsilon_1- \epsilon_2\over\hbar \omega_c} \right)^2 \right].
\end{eqnarray}
One sees that the correlation function exhibits harmonic oscillations with the period
$\hbar \omega_c$ up to the energies of the order of $\epsilon_1 - \epsilon_2 \sim
\hbar \omega_c \sqrt{\omega_c \tau_{tr}}$. The amplitude of these oscillation is
$\omega_c \tau_{tr} \gg 1$ times smaller than in the case of vertical tunneling into the
bulk of the two-dimensional electron gas, see Eq. (\ref{result}). 

\section{DOS correlation function in an interacting system}

Until now we have completely disregarded effects of the electron-electron interaction. It is
known, however, that this interaction has a crucial effect\cite{AAL} on the tunneling DOS of
the disordered conductor.  Namely, interaction leads to a strong energy dependence of the
single-particle  density of electron states for the energies close to the Fermi level. As a
result, the density of states must be written as a function depending both on the position of
the Fermi level and on the electron energy measured {\it from} the Fermi level:
\begin{equation}\label{nuint}
\nu (\epsilon ) = \nu (\epsilon - \epsilon_F, \epsilon_F).
\end{equation}
In the two-dimensional system  $\nu (\epsilon - \epsilon_F, \epsilon_F)$ has a
logarithmical singularity\cite{AAL} at small $\epsilon - \epsilon_F \ll \hbar /\tau_{tr}$, and
can be quite pronounced\cite{We1,We2} even at large $\epsilon - \epsilon_F \gg \hbar
/\tau_{tr}$. In particular, in the classical magnetic field $\nu (\epsilon - \epsilon_F,
\epsilon_F)$ is an oscillating function\cite{We2} of $\epsilon - \epsilon_F$ with a
characteristic period of cyclotron quantum $\hbar \omega_c$. 

As a consequence of Eq. (\ref{nuint}), the DOS correlation function for an interacting
system is a function of three arguments:
\begin{eqnarray}
P(\epsilon_1, \epsilon_2) & = & P (\epsilon_1 - \epsilon_F^{(1)}, \epsilon_2 -
\epsilon_F^{(2)},\epsilon_F^{(1)} - \epsilon_F^{(2)}) \nonumber \\
& = & {\left< \delta \nu (\epsilon_1 -
\epsilon_F^{(1)}, \epsilon_F^{(1)})\, \delta \nu (\epsilon_2 - \epsilon_F^{(2)},
\epsilon_F^{(2)}) \right> \over \nu_0^2}.
\end{eqnarray}

In order to observe experimentally the oscillations of the DOS correlation function
predicted in the present paper and given by Eqs.(\ref{result}) and (\ref{result22}), one has
to distinguish them from the {\it interaction}-induced oscillations of the density of the
electron states. The easiest way to do this is to fix two of the arguments of the correlation
function,
$\epsilon_1 -
\epsilon_F^{(1)}$ and
$\epsilon_2 - \epsilon_F^{(2)}$, and then measure $P$ as a function of the shift in the
chemical potential, $\epsilon_F^{(1)} - \epsilon_F^{(2)}$. 

\section{Conclusions}

In summary, we study properties of the two-dimentional  conductor
with a smooth disorder potential in a magnetic field. It is known that the average density of
states of such a conductor is hardly modified by the magnetic field [$\delta \nu /\nu_0
\propto \exp (-2\pi/\omega_c\tau_s)$] as long as
$\omega_c\tau_s\ll 1$. We show that despite such ``classical''  magnetic field does not
influence the average DOS of the conductor,  it  does  affect strongly the correlation
function of the local density of states,
$ P(\epsilon_1- \epsilon_2)$. Namely, provided that $\omega_c\tau_{tr}\gg 1$, the correlation
function $ P(\epsilon_1-\epsilon_2)$ aquires an oscillatory structure  with the
characteristic period $\hbar\omega_c$. This structure can be observed  in  
tunneling experiments on both vertical tunneling into the bulk of the two-dimensional
conductor, and lateral tunneling into the edge of the conductor. 

\acknowledgements{Support by NSF Grants DMR-94232444 and DMR-9731756, and by A.P. Sloan
Fellowship (I.L.A.) is gratefully acknowledged.}

\end{multicols}

\end{document}